\documentclass[11pt,a4wide]{article}

\usepackage{fullpage}
\usepackage{a4wide}
\usepackage{amsmath}
\usepackage{graphicx}
\usepackage{algorithm}
\usepackage{algorithmic}
\usepackage{natbib}
\usepackage{microtype}

\newcommand{\Matlab}{MATLAB\textsuperscript{\tiny\textregistered}}

\begin{document}



\title{Julia: A Fast Dynamic Language for Technical Computing}
\author{Jeff Bezanson\thanks{Email: {\tt bezanson@mit.edu}}\\MIT \and
  Stefan Karpinski\thanks{Email: {\tt stefan@karpinski.org}}\\MIT \and
  Viral B. Shah\thanks{Email: {\tt viral@mayin.org}} \and 
  Alan Edelman\thanks{email: {\tt edelman@math.mit.edu}}\\MIT}


\maketitle
\begin{abstract}
  Dynamic languages have become popular for scientific computing. They
  are generally considered highly productive, but lacking in performance.
  This paper presents Julia, a new dynamic language for technical
  computing, designed for performance from the beginning by adapting and
  extending modern programming language techniques. A design based on
  generic functions and a rich type system simultaneously enables an
  expressive programming model and successful type inference, leading to
  good performance for a wide range of programs. This makes it possible
  for much of Julia's library to be written in Julia itself, while also
  incorporating best-of-breed C and Fortran libraries.
\end{abstract}




\section{Introduction}

Convenience is winning. Despite advances in compiler technology and
execution for high-performance computing, programmers continue to prefer
high-level dynamic languages for algorithm development and data analysis
in applied math, engineering, and the sciences. High-level environments
such as \Matlab\!, Octave~\cite{Octave}, R~\cite{Rlang},
SciPy~\cite{numpy}, and SciLab~\cite{scilab} provide greatly increased
convenience and productivity. However, C and Fortran remain the gold
standard languages for computationally-intensive problems because
high-level dynamic languages still lack sufficient performance. As a
result, the most challenging areas of technical computing have benefited
the least from the increased abstraction and productivity offered by
higher level languages.

Two-tiered architectures have emerged as the standard compromise between
convenience and performance: programmers express high-level logic in a
dynamic language while the heavy lifting is done in C and Fortran. The
aforementioned dynamic technical computing environments are all themselves
instances of this design. While this approach is effective for some
applications, there are drawbacks. It would be preferable to write
compute-intensive code in a more productive language as well. This is
particularly true when developing parallel algorithms, where code
complexity can increase dramatically. Instead, there is pressure to write
``vectorized'' code, which is unnatural for many problems and might
generate large temporary objects which could be avoided with explicit
loops. Programming in two languages is more complex than using either
language by itself due to the need for mediation between different type
domains and memory management schemes. Interfacing between layers may add
significant overhead and makes whole-program optimization difficult.
Two-tiered systems also present a social barrier, preventing most users
from understanding or contributing to their internals.

An alternative to the two-tier compromise is to enhance the performance of
existing dynamic languages. There has been
significant progress along these lines. Projects like the Python
compiler framework PyPy \cite{pypyjit} have been fairly successful.
Similar efforts exist for languages from LISP onward. The common feature
of all such projects is that they seek to add performance to an existing
language. This is obviously useful, since existing code can benefit. While
promising, these efforts have yet to eliminate the need for the two-tier
approach in practice. Design decisions made under the assumption that
a language would be implemented as an interpreter tend to
sabotage the ability to generate efficient code. As Henry Baker observed
of Common LISP, ``...the polymorphic type complexity of the Common LISP
library functions is mostly gratuitous, and both the efficiency of
compiled code and the efficiency of the programmer could be increased by
rationalizing this complexity.''~\cite{nimble} Others have echoed these
sentiments~\cite{lispcrit}~\cite{evaluatingR}.

Julia is designed from the ground up to take advantage of modern
techniques for executing dynamic languages efficiently. As a result, Julia
has the performance of a statically compiled language while providing
interactive dynamic behavior and productivity like Python, LISP or Ruby.
The key ingredients of performance are:
\begin{itemize}
\item Rich type information, provided naturally by multiple dispatch;
\item Aggressive code specialization against run-time types;
\item JIT compilation using the LLVM compiler framework~\cite{LLVM}.
\end{itemize}
Although a sophisticated type system is made available to the programmer,
it remains unobtrusive in the sense that one is never required to specify
types. Type information flows naturally from having actual values (and
hence types) at the time of code generation, and from the language's core
paradigm: by expressing the behavior of functions using multiple dispatch,
the programmer unwittingly provides the compiler with extensive type
information.

We validate our design by implementing Julia's standard library, which
encompasses most of the core functionality of standard technical computing
environments, in Julia itself. As a result, our library code is more
generic and compact, and it is possible to inline library code into user
code and vice versa. Julia was announced as an open source project in
February 2012. New users have been easily able to read the standard
library code, modify it, imitate it, and extend it for their own purposes.
Our goals and work so far seem to have struck a chord --- a significant
community has grown around Julia in the short time since the initial
public announcement.

\section{Language Design}

Julia's primary means of abstraction is dynamic multiple dispatch.
Much of a language consists of mechanisms for selecting
code to run in different situations --- from method selection to
instruction selection. We use only dynamic multiple dispatch for this
purpose, which is possible through sufficiently expressive
dispatch rules.

To obtain the desired expressiveness and compile-time type information,
we must employ fairly sophisticated types. To balance this
requirement with usability, we have attempted to design the
``easy version'' of key features, namely parametric types and methods.
We provide these features without a concept of static application
(e.g. template instantiation), without distinct type and expression
contexts, and leaving parameters optional whenever possible.
Dynamic languages need parametric types so a compiler can keep track
of the types of values even when they are stored in shared mutable data
structures (optimizing compilers would need this feature even if the
types are not exposed in the language as they are in Julia).


Types may be used to make declarations, but we do not
require declarations for performance. To achieve this, Julia's compiler
automatically specializes methods for types encountered at run time
(or at compile time, to the extent types are known then). Effectively,
every method is a template (in the C++ sense) by default, with
parameterization and instantiation directed by the compiler.
We introduce some type-based heuristics for controlling method specialization.

\subsection{Rationale}

In past work on optimizing dynamic languages, researchers have observed that
programs are not as dynamic as their authors might think:
``We found that dynamic
features are pervasive throughout the benchmarks and the libraries they
include, but that most uses of these features are highly constrained...''
\cite{profileguided}. In this sense, the designs of existing dynamic languages
do not present a good trade-off. Much code is statically-typeable and could be
executed more efficiently, but the language designs and implementations do not
anticipate this fact.

We hypothesize that the following forms of ``dynamism'' are the most useful:

\begin{itemize}
\item The ability to run code at load time and compile time, eliminating
some of the distractions of build systems and configuration files.
\item A universal {\tt Any} type as the only true static type,
allowing the issue of static types to be ignored when desired.
\item Never rejecting code that is syntactically well-formed.
\item Behavior that depends only on run-time types (i.e. no static overloading).
\end{itemize}

Julia avoids some of the ``overly permissive'' features of systems like
CLOS \cite{clostrophobia} that get in the way of compiler optimizations,
using the following restrictions:

\begin{itemize}
\item Types themselves are immutable.
\item The type of a value cannot change over its lifetime.
\item Local variable environments are not reified.
\item Program code is immutable (but new code may be generated and executed at any time).
\item Not all bindings are mutable ({\tt const} identifiers are allowed).
\end{itemize}

These restrictions allow the compiler to see all uses of local
variables, and perform dataflow analysis on local variables using only
local information. This is important, since it allows user code to call
statically-unknown functions without interfering with optimizations done
around such call sites. Statically-unknown function calls arise in
many contexts, such as calling a function taken from an untyped data structure,
or dynamically dispatching a method call due to unknown argument types.

\subsection{Core Language Overview}

The core Julia language contains the following components:

\begin{enumerate}
\item A syntax layer, to translate surface syntax to a suitable
intermediate representation (IR).
\item A symbolic language and corresponding data structures for representing
certain kinds of types, and implementations of lattice operators ($meet$,
$join$, and $\leq$) for those types.
\item An implementation of generic functions and dynamic multiple dispatch
based on those types.
\item Compiler intrinsic functions for accessing the object model
(type definition, method definition, object allocation, element access,
testing object identity, and accessing type tags).
\item Compiler intrinsic functions for native arithmetic, bit string operations,
and calling native (C or Fortran) functions.
\item A mechanism for binding top-level names.
\end{enumerate}

The IR describes a function body as a sequence of assignment operations,
function calls, labels, and conditional branches. Julia's semantics
are those of a standard imperative language: statements are executed in order,
with function arguments evaluated eagerly. All values are conceptually
references, and are passed by reference as in LISP.

\subsection{Types}

Julia treats types as symbolic descriptions of sets of values. Every value has
a unique, immutable, run-time implementation type. Objects carry type tags, and
types themselves are Julia objects that can be created and inspected
at run time.
Julia has five kinds of types:

\begin{enumerate}
\item Abstract types, which may have declared subtypes and supertypes
      (a subtype relation is declared using the notation {\tt Sub~<:~Super})
\item Composite types (similar to C structs\footnote{Currently, composite
types are mutable, but we plan to make mutability optional.}), which have
named fields and declared supertypes
\item Bits types, whose values are represented as bit strings, and which
      have declared supertypes
\item Tuples, immutable ordered collections of values
\item Union types, abstract types constructed from other types via set union
\end{enumerate}

Bits types allow users to add new fixed-width number-like types and obtain the
same performance that primitive numeric types enjoy in other systems. Julia's
``built in'' numeric types are defined as bits types. Julia method dispatch
is based on types rather than field lookup, so whether a value is of a bits
type or composite type is a representation detail that is generally
invisible.

Tuples are used to represent the types of method arguments and multiple
return values. The type of a tuple is
defined recursively as a tuple of the types of its elements. Tuple types are
covariant; a tuple type is a subtype of another if
its elements are subtypes of the corresponding elements of the other.
Tuple types may end in a special {\tt ...} type that
indicates any number of elements may be added. This is used to express the
types of variadic methods. For example the type {\tt (String, Int...)}
indicates a tuple where the first element is a String and any number of
trailing integers may be present.

Union types are used primarily to construct tight least upper bounds
when the inference algorithm needs to join unrelated types. For example,
a method might return an {\tt Int} or a {\tt String} in separate
arms of a conditional. In this case its type can be inferred as
{\tt Union(Int,String)}. Union types are also useful for defining
ad-hoc type hierarchies different from those imagined when the types
involved were first defined. Lastly,
union types can be used to declare methods applicable to multiple types.

\subsection{Type Parameters}

Abstract types, composite types, and bits types may have parameters, which
makes it possible to express variants of a given type (for example, array types
with different element types). These types are all invariant with respect to
their parameters (i.e. two versions of the same type with different parameters
are simply different, and have no subtype relationship). Type
constructors are applied using curly braces, as in {\tt Array\{Float64,1\}}
(the {\tt Array} type is parameterized by element type and rank).
Semantically, a type constructor application is a function call expression
evaluated at run time.

Type parameters may have bounds \cite{boundedquant}, also declared using
the {\tt <:} operator, as in {\tt Rational\{T<:Integer\}}.

To help meet our goal of convenience, we allow writing
parametric types without parameters, or omitting trailing parameters.
{\tt Array} refers to any kind of dense array, and {\tt Array\{Float64\}}
refers to a Float64 Array of any rank.
The result of one of these expressions is effectively an ad-hoc abstract
supertype of all instantiations one could obtain by filling in the missing
parameters.

This design also makes it easy to add parameters to types later; existing
code does not need to be modified.

\subsection{Generic Functions}

The vast majority of Julia functions (in both the library and user programs)
are generic functions, meaning they contain multiple definitions or methods for
various combinations of argument types. When a generic function is applied,
the most specific definition that matches the run-time argument types is
invoked. Generic functions have appeared in several object systems in the past,
notably CLOS \cite{closoverview} and Dylan \cite{dylanlang}.
Julia is distinguished from these in that
it uses generic functions as its primary abstraction mechanism, putting it in
the company of research languages like Diesel \cite{dieselspec} and Cecil
\cite{cecil}. Aside
from being practical for mathematical styles of programming,
this design is satisfying also because it permits
expression of most of the popular patterns of object-oriented programming,
while leaving the core language with fewer distinct features.

\subsection{Method Definition}
Method definitions have a long (multi-line) form and a short form.

\begin{verbatim}
function iszero(x::Number)
    return x==0
end

iszero(x) = (x==0)
\end{verbatim}

A type declaration with {\tt ::} on an argument is a dispatch specification.
When types are omitted, the default is {\tt Any}.
A {\tt ::} expression may be added to any program expression, in which case
it acts as a run-time type assertion. As a special case, when {\tt ::} is
applied to a variable name in statement position (a construct which otherwise
has no effect) it means the variable \emph{always} has the specified type,
and values will be converted to that type (by calling {\tt convert}) on
assignment to the variable.

Note that there is no distinct type context; types are computed by ordinary
expressions
evaluated at run time. For example, {\tt f(x)::Int} is lowered to the
function call {\tt typeassert(f(x),Int)}.

Anonymous functions are written using the syntax {\tt x->x+1}.

Local variables are introduced implicitly by assignment. Modifying a
global variable requires a {\tt global} declaration.

Operators are simply functions with special calling syntax. Their
definitions look the same as those of ordinary functions, for example
{\tt +(x,y)~=~...}, or {\tt function~+(x,y)}.

When the last argument in a method signature is followed by {\tt ...}
the method accepts any number of arguments, and the last argument name
is bound to a tuple containing the tail of the argument list. The syntax
{\tt f(t...)} ``splices'' the contents of an iterable object {\tt t} as the
arguments to {\tt f}.

Generic functions are a natural fit for mathematical programming. For example,
consider implementing exponentiation (the {\tt \^{}} operator in Julia).
This function
lends itself to multiple definitions, specializing on both arguments
separately: there might be one definition for two floating-point numbers that
calls a standard math library routine, one definition for the case where the
second argument is an integer, and separate definitions for the case where the
first argument is a matrix. In Julia these signatures would be written as
follows:

\begin{verbatim}
function ^(x::Float64, p::Float64)
function ^(x, p::Int)
function ^(x::Matrix, p)
\end{verbatim}

\subsection{Parametric Methods}

It is often useful to refer to parameters of argument types inside methods,
and to specify constraints on those parameters for dispatch purposes.
Method parameters address these needs. These parameters behave a bit like
arguments, but they are always derived automatically from
method argument types and not specified explicitly by the caller.
The following signature presents a typical example:

\begin{verbatim}
function assign{T<:Integer}(a::Array{T,1}, i, n::T)
\end{verbatim}

This signature is applicable to 1-dimensional arrays whose element type is
some kind of integer, any type of second argument, and a third argument
that is the same type as the array's element type. Inside the method,
{\tt T} will be bound to the array element type.

The primary use of this construct is to write methods applicable to a
family of parametric types
(e.g. all integer arrays, or all numeric arrays)
despite invariance. The other use is
writing ``diagonal'' constraints as in the example above. Such diagonal
constraints significantly complicate the type lattice operators.

\subsection{Constructors}

Composite types are applied as functions to construct instances.
The default constructor accepts values for each field as arguments.
Users may override the default constructor by writing method definitions
with the same name as the type inside the type definition block. Inside the
{\tt type} block the identifier {\tt new} is bound to a pseudofunction
that actually constructs instances from field values. The constructor
for the {\tt Rational} type is a good example:

\begin{verbatim}
type Rational{T<:Integer} <: Real
    num::T
    den::T

    function Rational(num::T, den::T)
        if num == 0 && den == 0
            error("invalid rational: 0//0")
        end
        g = gcd(den, num)
        new(div(num, g), div(den, g))
    end
end
\end{verbatim}

This allows {\tt Rational} to enforce representation as a fraction in
lowest terms.

\subsection{Singleton Kinds}

A generic function's method table is effectively a dictionary where the keys
are types. This suggests that it should be just as easy to define or look up
methods with types themselves as with the types of values. Defining methods on
types directly is analogous to defining class methods in class-based object
systems. With multi-methods, definitions can be associated with combinations
of types, making it easy to represent properties not naturally owned by one
type.

To accomplish this, we introduce a special singleton kind {\tt Type\{T\}},
which contains the type {\tt T} as its only value.
The result is a feature similar to {\tt eql}
specializers in CLOS, except only for types. An example use is defining
type traits:

\begin{verbatim}
typemax(::Type{Int64}) = 9223372036854775807
\end{verbatim}

This definition will be invoked by the call {\tt typemax(Int64)}. Note that
the name of a method argument can be omitted if it is not referenced.

Types are useful as method arguments in several other cases. One example is
file I/O, where a type can be used to specify what to read. The call
{\tt read(file,Int32)} reads a 4-byte integer and returns it as an {\tt Int32}
(a fact that the type inference process is able to discover). We find this
more elegant and convenient than systems where enums or special constants must
be used for this purpose, or where the type information is implicit
(e.g. through return-type overloading).

This feature allows sharper types to be inferred when the user programs
with types, for example by calling the {\tt typeof} function or applying
type constructors. As a result, we gain the performance and flexibility
advantages of static parameters (such as template arguments) without special
syntax.

\subsection{Method Sorting and Ambiguity}
Methods are stored sorted by specificity, so the first matching method
(as determined by the subtype predicate) is always the correct one to invoke.
This means much of the dispatch logic is contained in the sorting process.
Comparing method signatures for specificity is not trivial. As one might
expect, the ``more specific''\footnote{Actually, ``not less specific'',
since specificity is a partial order.}
predicate is quite similar to the subtype
predicate, since a type that is a subtype of another is indeed more specific
than it. However, a few additional rules are necessary to capture the
intuitive concept of ``more specific''. Our formal definition is
summarized as the disjunction of the following rules ($A$ is more specific
than $B$ if):

\begin{enumerate}
\item $A$ is a subtype of $B$
\item $A$ is of the form {\tt T\{P\}} and $B$ is of the form {\tt S\{Q\}}, and
$T$ is a subtype of $S$ for some parameter values
\item The intersection of $A$ and $B$ is nonempty, more specific than $B$, and
not equal to $B$, and $B$ is not more specific than $A$
\item $A$ and $B$ are tuple types, $A$ ends in a vararg ({\tt ...}) type,
and $A$ would be more specific than $B$ if its vararg type were expanded to
give it the same number of elements as $B$
\item $A$ and $B$ have parameters and compatible structures, and $A$
provides a consistent assignment for $B$'s parameters, but not the other
way around
\end{enumerate}

Rule 2 means that declared subtypes are always more specific than their
declared supertypes regardless of type parameters. Rule 3 is mostly useful for
union types: if $A$ is {\tt Union(Int32,String)} and $B$ is {\tt Number}, $A$
should
be more specific than $B$ because their intersection ({\tt Int32}) is clearly
more specific than $B$. Rule 4 means that argument types are more important for
specificity than argument count; if $A$ is {\tt (Int32...)} and $B$ is
{\tt (Number, Number)} then $A$ is more specific.

Rule 5 makes diagonal constraints more specific; $\forall T (T,T)$ is more
specific than $\forall X,Y (X,Y)$.
The specificity of a type variable is determined extensionally, i.e.
according to the set of values it would ultimately encompass. For example,
{\tt T~<:~Number} has the same specificity as {\tt Number}. This approach
has been found useful in past work combining parametric types and multiple
dispatch \cite{modularmultipledispatch}.

Julia uses symmetric multiple dispatch, which means all arguments
are equally important. Therefore, ambiguous signatures are possible.
For example, given {\tt foo(x::Number, y::Int)} and
{\tt foo(x::Int, y::Number)} it is not clear which method to call when both
arguments are integers. We detect ambiguities when a method is added, by
looking for a pair of signatures with a non-empty intersection where neither
one is more specific than the other. A warning message is displayed for each
ambiguity, showing the user the computed type intersection so it is clear what
definition is missing. For example:

\begin{verbatim}
Warning: New definition foo(Int,Number) is
   ambiguous with foo(Number,Int). Make sure
   foo(Int,Int) is defined first.
\end{verbatim}

\subsection{Iteration}

A {\tt for} loop is translated to a while loop with method calls according
to an iteration interface ({\tt start}, {\tt done}, and {\tt next}).

\begin{verbatim}
for i in range
    # body
end
\end{verbatim}

Becomes:

\begin{verbatim}
state = start(range)
while !done(range, state)
  (i, state) = next(range, state)
  # body
end
\end{verbatim}

This design for iteration was chosen because it is not tied to mutable
heap-allocated state, such as an iterator object that updates itself.

\subsection{Special Operators}

Special syntax is provided for certain functions.

\begin{tabular}{|l|l|}\hline
surface syntax     & lowered form \\\hline \hline
{\tt a[i, j]}      & {\tt ref(a, i, j)} \\\hline
{\tt a[i, j] = x}  & {\tt assign(a, x, i, j)} \\\hline
{\tt [a; b]}       & {\tt vcat(a, b)} \\\hline
{\tt [a, b]}       & {\tt vcat(a, b)} \\\hline
{\tt [a b]}        & {\tt hcat(a, b)} \\\hline
{\tt [a b; c d]}   & {\tt hvcat((2,2), a, b, c, d)}\\\hline
\end{tabular}

\subsection{Calling C and Fortran}

We provide the keyword {\tt ccall} for calling native code in-line.
Its syntax looks like a function call, where the programmer specifies
an address, result and argument types, and argument values:

\begin{verbatim}
ccall(dlsym(libm, :sin), Float64, (Float64,), x)
\end{verbatim}

The first three arguments to {\tt ccall} are actually pseudo-arguments,
evaluated at compile time. The compiler front-end inserts calls to the
{\tt convert} function for each argument, ensuring that the actual arguments
will match the provided signature.

In Fortran, all arguments are passed by reference. To handle this,
argument types must be written as pointer types such as
{\tt Ptr\{Float64\}}. An ambiguity then arises in argument conversion:
an integer argument could be interpreted as a pointer, or as a number
to convert to {\tt Float64} and pass by reference. To resolve this
ambiguity, the programmer can request the second interpretation by
prefixing an argument with an ampersand (a pun on C syntax for taking the
address of a value), as in {\tt \&x}.

\subsection{Parallelism}

Parallel execution is provided by a message-based multi-processing
system implemented in Julia in the standard library.
The language design supports the implementation of such libraries by
providing symmetric coroutines, which can also be thought of as
cooperatively scheduled threads. This feature allows asynchronous
communication to be hidden inside libraries, rather than requiring the
user to set up callbacks. Julia does not currently support
native threads, which is a limitation, but has the advantage of avoiding
the complexities of synchronized use of shared memory.

\subsection{Design Limitations}

In our design, type information always flows along with values, in the
forward control flow direction. This prevents us from doing certain tricks
that static type systems are capable of, such as return-type overloading.
Return-type overloading requires a robust notion of the type of a value
\emph{context}---the type expected or required of some term---in order to
select code on that basis. There are other cases where ``backwards'' type
flow might be desirable, such as determining the type of a container based
on the type of a value stored into it at a later program point. It may be
possible to get around this limitation in the future using inversion of
control---passing a function argument whose result type has already been
inferred, and using that type to construct a container before elements are
computed.

Modularity is a perennial difficulty with multiple dispatch, as any
function might apply to any type, and there is no point where functions or
types are closed to future definitions. Thus at the moment Julia is
essentially a whole-program compiler. We plan to implement a module system
that will at least allow code to control which name bindings and definitions
it sees. Such modules could be separately compiled to the extent that
programmers are willing to ask for their definitions to be ``closed''.

Lastly, at this time Julia uses a bit more memory than we would prefer.
Our compiler data structures, type information, and generated native code
take up more space than the compact bytecode representations used by many
dynamic languages.

\section{Implementation}

Much of the implementation is organized around method dispatch. The dispatch
logic is both a large portion of the behavior of Julia functions, and the
entry point of the compiler's type inference and specialization logic.

\subsection{Method Caching and Specialization}

The first step of method dispatch is to look for the argument types in a
per-function cache. The cache has an entry for (almost) every set of concrete
types to which the function has been applied. Concrete types are hash-consed,
so they can be compared by simple pointer comparison. This makes cache lookup
faster than the $subtype$ predicate. As part of hash-consing, concrete types
are assigned small integer IDs. The ID of the first argument is used as a
primary key into a method cache, so when signatures differ only in the
type of the first argument a simple indexed lookup suffices.

On a cache miss, a slower search for the matching definition is performed using
$subtype$.
Then, type inference is invoked on the matching method using the types
of the actual arguments. The resulting type-annotated and optimized method is
stored in the cache. In this way, method dispatch is the primary source of type
information for the compiler.

\subsection{Method Specialization Heuristics}

Our aggressive use of code specialization has the obvious pitfall that it might
lead to excessive code generation, consuming memory and compile time. We found
that a few mild heuristics suffice to give a usable system with reasonable
resource requirements.

The first order of business is to ensure that the dispatch and specialization
process converges. The reason it might not is that our type inference algorithm
is implemented in Julia itself. Calling a method on a certain type $A$ can cause
the type inference code to call the same method on type $B$, where types
$A$ and $B$
follow an infinite ascending chain in either of two partial orders (the
$typeof$ order or the $subtype$ order). Singleton kinds are the most
prominent example, as type inference might attempt to successively consider
{\tt Int32}, {\tt Type\{Int32\}}, {\tt Type\{Type\{Int32\}\}}, and so on. We
stop this process by replacing any nestings of {\tt Type} with the
unspecialized version of {\tt Type} during method specialization (unless the
original method declaration actually specified a type like
{\tt Type\{Type\{Int32\}\}}).

The next heuristic avoids specializing methods for tuple types of every length.
Tuple types are cached as the intersection of the declared type of the method
slot with the generic tuple type {\tt (Any...)}. This makes the resulting cache
entry valid for any tuple argument, again unless the method declaration
contained a more specific tuple type. Note that all of these heuristics require
corresponding changes in the method cache lookup procedure, since they yield
cache entries that do not have to exactly match candidate arguments.

A similar heuristic is applied to variadic methods, where we wish to avoid
caching argument lists of every length. This is done by capping argument lists
at the length of the longest signature of any method in the same generic
function. The ``capping'' involves replacing the last argument with a
{\tt ...} type. Ideally, we want to form the biggest type that's not a
supertype of any other method signatures. However, this is not always possible
and the capped type might conflict with another signature. To deal with this
case, we find all non-empty intersections of the capped type with other
signatures, and add dummy cache entries for them. Hitting one of these entries
alerts the system that the arguments under consideration are not really in the
cache. Without the dummy entries, some arguments might incorrectly match the
capped type, causing the wrong method to be invoked.

The next heuristic concerns singleton kinds again. Because of the singleton
kind feature, every distinct type object ({\tt Any}, {\tt Number}, {\tt Int},
etc.) passed to a method might trigger a new specialization. However, most
methods are not ``class methods'' and are not concerned with type objects.
Therefore, if no method definition in a certain function involves {\tt Type}
for a certain argument slot, then that slot is not specialized for different
type objects.

Finally, we introduce a special type {\tt ANY} that can be used in a method
signature to hint that a slot should not be specialized. This is used in the
standard library in a small handful of places, and in practice is less
important than the heuristics described above.

\subsection{Type Inference}

Types of program expressions and variables are inferred by forward
dataflow analysis\footnote{Adding a reverse dataflow pass could potentially
improve type information, but we have not yet done this.}.
The original algorithm for such dynamic type inference was given by
Kaplan and Ullman \cite{kaplanullman}.
This is different from type inference in the ML family of languages
\cite{MLtypeinf}, where the compiler \emph{must} be able to determine
types, using an algorithm based on unification. Dynamic type inference
has been applied to LISP \cite{TICL} \cite{pticl} \cite{nimble},
and object-oriented languages such as Self \cite{selflang} and
JavaScript \cite{typeinfjavascript}.

We determine a maximum fixed-point (MFP) solution using
Algorithm~\ref{alg1}, based on
Mohnen's graph-free dataflow analysis framework \cite{graphfree}. The basic
idea is to keep track of the state (the types of all variables) at each program
point, determine the effect of each statement on the state, and ensure that
type information from each statement eventually propagates to all other
statements reachable by control flow. We augment the basic algorithm with
support for mutually-recursive functions
(functions are treated as program points that might need to be revisited).


The origin of the type information used by the MFP algorithm is
evaluation of known functions over the type domain \cite{abstractinterp}.
This is done by the $eval$ subroutine. The $interpret$ subroutine calls
$eval$, and also handles assignment statements by returning the new types
of affected variables. Each known function
call is either to one of the small number of built-in functions, in which
case the result type is computed by a (usually trivial) hand-written
type transfer function, or to a generic function, in which case the result
type is computed by recursively invoking type inference. In the generic
function case, the inferred argument types are met ($\sqcap$) with the
signatures of each method definition. Matching methods are those where the
meet (greatest lower bound)
is not equal to the bottom type ({\tt None} in Julia).
Type inference is invoked on each matching
method, and the results are joined ($\sqcup$) together. The following equation
summarizes this process:

\[
T(f,t_{arg}) = \bigsqcup_{(s,g) \in f}T(g,t_{arg} \sqcap s)
\]

\noindent
$T$ is the type inference function.
$t_{arg}$ is the inferred argument tuple type. The tuples $(s,g)$
represent the signatures $s$ and their associated definitions $g$ within
generic function $f$.

Two optimizations are helpful here. First, it is rarely
necessary to consider all method definitions. Since methods are stored in
sorted order, as soon as the union of the signatures considered so far is a
supertype of $t_{arg}$, no more definitions need to be considered.
Second, the join operator employs \emph{widening} \cite{widening}:
if a type becomes too large it may simply return {\tt Any}. In this case
the recursive inference process may stop immediately.

\begin{algorithm}
\caption{Infer function return type}
\label{alg1}
\begin{algorithmic}
\REQUIRE function $F$, argument type tuple $A$, abstract execution stack $S$
\ENSURE result type $S.R$
\STATE $V \leftarrow$ set of all locally-bound names
\STATE $V_{a} \leftarrow$ argument names
\STATE $n \leftarrow length(F)$
\STATE $W \leftarrow \{1\}$ \COMMENT {set of program counters}
\STATE $P_r \leftarrow \emptyset$ \COMMENT {statements that recur}
\STATE $\forall v \in V, \Gamma[1][v] \leftarrow \text{Undef}$
\STATE $\forall i, \Gamma[1][V_{a}[i]] \leftarrow A[i]$  \COMMENT {type environment for statement 1}
\WHILE{$W \neq \emptyset$}
 \STATE $p \leftarrow \operatorname{choose}(W)$
 \REPEAT
  \STATE $W \leftarrow W - p$
  \STATE $new \leftarrow interpret(F[p],\Gamma[p],S)$
  \IF {$S.rec$}
   \STATE $P_r \leftarrow P_r \cup \{p\}$
   \STATE $S.rec \leftarrow \text{false}$
  \ENDIF
  \STATE $p^{\prime} \leftarrow p+1$
  \IF{$F[p] = $\texttt{(goto l)}}
   \STATE $p^{\prime} \leftarrow l$
  \ELSIF{$F[p] = $\texttt{(gotoif cond l)}}
   \IF {\NOT $new \leq \Gamma[l]$}
    \STATE $W \leftarrow W \cup \{l\}$
    \STATE $\Gamma[l] \leftarrow \Gamma[l] \sqcup new$
   \ENDIF
  \ELSIF{$F[p] = $\texttt{(return e)}}
   \STATE $p^{\prime} \leftarrow n+1$
   \STATE $r \leftarrow eval(e,\Gamma[p],S)$
   \IF {\NOT $r \leq S.R$}
    \STATE $S.R \leftarrow S.R \sqcup r$
    \STATE $W \leftarrow W \cup P_r$
   \ENDIF
  \ENDIF
  \IF{$p^{\prime} \leq n$ \AND \NOT $new \leq \Gamma[p^{\prime}]$}
   \STATE $\Gamma[p^{\prime}] \leftarrow \Gamma[p^{\prime}] \sqcup new$
   \STATE $p \leftarrow p^{\prime}$
  \ENDIF
 \UNTIL{$p^{\prime} = n+1$}
\ENDWHILE
\STATE {$S.rec \leftarrow P_r \neq \emptyset$}
\end{algorithmic}
\end{algorithm}

\subsubsection{Interprocedural Type Inference}

Type inference is invoked through ``driver'' Algorithm~\ref{alg2}
which manages mutual recursion and memoization of inference results.
A stack of abstract activation records is maintained and used to detect
recursion. Each function has a property $incomplete(F,A)$ indicating that
it needs to be revisited when new information is discovered about the
result types of functions it calls. The $incomplete$ flags collectively
represent a set analogous to $W$ in Algorithm~\ref{alg1}.

The outer loop in Algorithm~\ref{alg2} looks for an existing activation
record for its input function and argument types. If one is found, it
marks all records from that point to the top of the stack, identifying
all functions involved in the call cycle. These marks
are discovered in Algorithm~\ref{alg1} when $interpret$ returns, and all
affected functions are considered $incomplete$. Algorithm~\ref{alg2}
continues to re-run inference on incomplete functions, updating the
inferred result type, until no recursion occurs or the result type
converges.

\begin{algorithm}
\caption{Interprocedural type inference}
\label{alg2}
\begin{algorithmic}
\REQUIRE function $F$, argument type tuple $A$, abstract execution stack $S$
\ENSURE returned result type
\STATE $R \leftarrow \bot$
\IF {$recall(F,A)$ exists}
 \STATE $R \leftarrow recall(F,A)$
 \IF {\NOT $incomplete(F,A)$}
  \RETURN $R$
 \ENDIF
\ENDIF
\STATE $f \leftarrow S$
\WHILE {\NOT $\operatorname{empty}(f)$}
 \IF {$f.F$ is $F$ \AND $f.A=A$}
  \STATE $r \leftarrow S$
  \WHILE {\NOT $r=tail(f)$}
   \STATE $r.rec \leftarrow \text{true}$
   \STATE $r \leftarrow tail(r)$
  \ENDWHILE
  \RETURN $f.R$
 \ENDIF
 \STATE $f \leftarrow tail(f)$
\ENDWHILE
\STATE $S^{\prime} \leftarrow extend(S, Frame(F,A,R,rec=\text{false}))$
\STATE invoke Algorithm~\ref{alg1} on $F,A,S^{\prime}$
\STATE $recall(F,A) \leftarrow S^{\prime}.R$
\STATE $incomplete(F,A) \leftarrow (S^{\prime}.rec \land \neg(R=S^{\prime}.R))$
\RETURN $S^{\prime}.R$
\end{algorithmic}
\end{algorithm}


\subsection{Lattice Operators}

Our type lattice is complicated by the presence of type parameters, unions,
and diagonal type constraints in method signatures. Fortunately, for our
purposes only the $\leq$ ($subtype$) relation needs to be computed accurately,
as it bears final responsibility for whether a method is applicable to
given arguments. Type union and intersection, used to estimate
least upper bounds and greatest lower bounds, respectively, may both be
conservatively approximated. If their results are too coarse, the
worst that can happen is performing method dispatch or type checks
at run time, since the inference process will simply conclude that it does
not know precise types.

A complication arises from the fact that our abstract domain is
available in a first-class fashion to user programs. When a program
contains a type-valued expression, we want to know which type it will
evaluate to, but this is not possible in general. Therefore in addition
to the usual \emph{type imprecision} (not knowing the type of a value),
we must also model \emph{type uncertainty}, where a type itself is
known imprecisely. A common example is application of the {\tt typeof}
primitive to a value of imprecise type. What is the abstract result of
{\tt typeof(x::Number)}? We handle this using bounded type variables,
effectively representing a \emph{range} rather than a point within the
type lattice. In this example, the
transfer function for {\tt typeof} is allowed to return
{\tt Type\{T<:Number\}}, where {\tt T} is a new type variable.

\subsubsection{Subtype Predicate}

See Algorithm~\ref{alg3}. Note that extensional type equality can be
computed as $(A\leq~B\land~B\leq~A)$, and this is used for types in
invariant context (i.e. type parameters). The algorithm uses subroutines
$p(A)$ which gives the parameters of type $A$, and $super(A)$ which gives
the declared supertype of $A$.

\begin{algorithm}
\caption{Subtype}
\label{alg3}
\begin{algorithmic}
\REQUIRE types $A$ and $B$
\ENSURE $A \leq B$
\IF {$A$ is a tuple type}
 \IF {$B$ is not a tuple type}
  \RETURN false
 \ENDIF
 \FOR {$i=1$ \TO $length(A)$}
  \IF {$A[i]$ is $T...$}
   \IF {$last(B)$ exists and is not $S...$}
    \RETURN false
   \ENDIF
   \RETURN $subtype(T,B[j])), i \leq j \leq length(B)$
  \ELSIF {$i > length(B)$ \OR \NOT $subtype(A[i],B[i])$}
   \RETURN false
  \ELSIF {$B[i]$ is $T...$}
   \RETURN $subtype(A[j],T)), i < j \leq length(A)$
  \ENDIF
 \ENDFOR
\ELSIF {$A$ is a union type}
 \RETURN $\forall t \in A, subtype(t,B)$
\ELSIF {$B$ is a union type}
 \RETURN $\exists t \in B, subtype(A,t)$
\ENDIF
\WHILE {$A \neq \texttt{Any}$}
 \IF {$typename(A) = typename(B)$}
  \RETURN {$subtype(p(A),p(B)) \land subtype(p(B),p(A))$}
 \ENDIF
 \STATE $A \leftarrow super(A)$
\ENDWHILE
\IF {$A$ is of the form {\tt Type\{T\}}}
 \RETURN $subtype(typeof(p(A)[1]),B)$
\ELSIF {$B$ is of the form {\tt Type\{T\}}}
 \STATE $B \leftarrow p(B)[1]$
 \RETURN $subtype(A,B) \land subtype(B,A)$
\ENDIF
\RETURN $B = \texttt{Any}$
\end{algorithmic}
\end{algorithm}

\subsubsection{Type Union}

Since our type system directly supports unions, the union of $T$ and
$S$ can be computed simply by constructing the type {\tt Union(T,S)}.
An obvious simplification is performed: if one of $T$ or $S$ is a
subtype of the other, it can be removed from the union. Nested union
types are flattened, followed by pairwise simplification.

\subsubsection{Type Intersection}

This is the difficult one: given types $T$ and $S$, we must try to compute
the smallest type $R$ such that
$\forall s, s \in T \land s \in S \Rightarrow s \in R$.
The conservative solution is to give up on finding the smallest such type, and
return \emph{some} type with this property. Simply returning $T$ or $S$
suffices for correctness, but in practice this algorithm
makes the type inference process nearly useless. A slightly better
algorithm is to check whether one argument is a subtype of the other, and
return the smaller type. It is also possible to determine quickly, in
many cases, that two types are disjoint, and return $\bot$. With these
two enhancements we start to obtain some useful type information. However,
we need to do better to take full advantage of the framework set up
so far.

Our algorithm has two phases. First, the structures of the two input types
are analyzed in a manner similar to $subtype$, except a constraint
environment is built, with entries $T\leq S$ for type variables $T$ in
covariant contexts (tuples) and entries $T=S$ for type variables $T$ in
invariant contexts (type parameters). In the second phase the constraints
are solved with an algorithm similar to that
used by traditional polymorphic type systems \cite{MLtypeinf}.

The code for handling tuples and union types is similar to that in
Algorithm~\ref{alg3}, so we focus instead on intersecting types in the
nominal hierarchy (Algorithm~\ref{alg4}). The base case occurs when
the input types are from the same family, i.e. have the same
$typename$. All we need to do is visit each parameter to collect any
needed constraints, and otherwise check that the parameters are equal.
When a parameter is a type variable, it is effectively covariant, and
must be intersected with the corresponding parameter of the other type
to form the final result.

\begin{algorithm}
\caption{Intersection of nominal types}
\label{alg4}
\begin{algorithmic}
\REQUIRE types $A$ and $B$, current constraint environment
\ENSURE return $T$ such that $A \sqcap B \leq T$, updated environment
\IF {$typename(A) = typename(B)$}
 \STATE $pa \leftarrow \operatorname{copy}(p(A))$
 \FOR {$i=1$ \TO $length(p(A))$ }
  \IF {$p(A)[i]$ is a typevar}
   \STATE {add $(p(A)[i]=p(B)[i])$ to constraints}
  \ELSIF {$p(B)[i]$ is a typevar}
   \STATE {add $(p(B)[i]=p(A)[i])$ to constraints}
  \ENDIF
  \STATE $pa[i] \leftarrow intersect(p(A)[i],p(B)[i])$
 \ENDFOR
 \RETURN {$typename(A)\{pa...\}$}
\ELSE
 \STATE $sup \leftarrow intersect(super(A),B)$
 \IF {$sup = \bot$}
  \STATE $sup \leftarrow intersect(A,super(B))$
  \IF {$sup = \bot$}
   \RETURN $\bot$
  \ELSE
   \STATE $sub \leftarrow B$
  \ENDIF
 \ELSE
  \STATE $sub \leftarrow A$
 \ENDIF
 \STATE $E \leftarrow conform(sup, super\_decl(sub))$
 \IF {$E$ contains parameters not in $formals(sub)$}
  \RETURN $\bot$
 \ENDIF
 \RETURN $intersect(sub, typename(sub)\{E...\})$
\ENDIF
\end{algorithmic}
\end{algorithm}

When the argument types are not from the same family, we recur up the
type hierarchy to see if any supertype of one of the arguments matches
the other. If so, the recursion gives us the intersected supertype $sup$,
and we face the problem of mapping it to the family of the original argument
type. To do this, we first call subroutine $conform$, which takes two types
with the same structure and returns an environment $E$ mapping any
type variables in one to their corresponding components in the other.
$super\_decl(t)$ returns the type template used by $t$ to instantiate its
supertype. If all goes well, this tells us what parameters $sub$ would
have to be instantiated with to have supertype $sup$. If, however, $E$
contains type variables not controlled by $sub$, then there is no way
a type like $sub$ could have the required supertype, and the overall answer
is $\bot$.
Finally, we apply the base case to intersect $sub$ with the type obtained
by instantiating its family with parameter values in $E$.

We use a simple algorithm to solve the type parameter constraints.
Constraints $T\leq S$ where $S$ is a concrete type are converted to
$T=S$ to help sharpen the result type.
If there are any conflicting constraints ($T=S$ and $T=U$ where $S\neq U$),
the type intersection is empty. If each type variable has exactly one
constraint $T=U$, we can substitute $find(X,U)$ for each occurrence
of $T$ in the computed type intersection, and we have a final answer.
$find$ works in the \emph{union-find} sense, following chains of equalities
until we hit a non-variable or an unconstrained variable. Unconstrained
type variables may be left in place.

The remaining case is type variables with multiple constraints. Finding
a satisfying assignment requires intersecting all the upper bounds for
a variable. It is here that we choose to throw in the towel and switch
to a coarser notion of intersection, denoted by $\sqcap^{*}$.
Intersection is effectively the inner loop of type inference, so in the
interest of getting a reasonable answer quickly we might pick
$X\sqcap^{*}Y=X$. A few simple heuristics might as well be added; for
example cases like two non-parameterized types where one is an immediate
subtype of the other can be supported easily.

In our implementation, type intersection handles most of the
complexity surrounding type variables and parametric methods.
It is used to test applicability of parametric methods; since all
run-time argument lists are of concrete type, intersecting their types
with method signatures behaves like $subtype$, except static parameters
are also properly matched. If intersection returns $\bot$ or does not find
values for all static parameters for a method, the method is not applicable.
Therefore in practice we do not really have the freedom to implement
$\sqcap$ and $\sqcap^{*}$ any way that obeys our correctness property.
They must be at least as accurate as $subtype$ in the case where one
argument is concrete.

\subsubsection{Widening Operators}

Lattices used in practical program analyses often fail to obey the finite
chain condition necessary for the MFP algorithm to converge (i.e. they
are not of finite height) and ours is no exception.

Widening is applied in two places: by the join operator, and on every
recursive invocation of type inference.  When a union type becomes too
large (as determined by a cutoff), it is replaced with {\tt
  Any}. Tuple types lend themselves to two infinite chains: one in
depth ({\tt (Any,)}, {\tt ((Any,),)}, {\tt (((Any,),),)}, etc.)  and
one in length ({\tt (Any...,)}, {\tt (Any,Any...,)}, {\tt
  (Any,Any,Any...,)}, etc.). These chains are capped at arbitrary
cutoffs each time the inference process needs to construct a tuple
type.

\subsection{Code Generation and Optimization}

After type inference is complete, we annotate each expression with its
inferred type. We then run two key optimization passes.
If the inferred argument types in a method call
indicate that a single method matches, we are free to inline that method.
For methods that return multiple values, inlining often yields
expressions that construct tuples and immediately take them apart. The
next optimization pass identifies these cases and removes the tuple
allocations.

The next set of optimizations is applied during code generation.
Our code generator targets the LLVM compiler framework \cite{LLVM}.
First, we examine uses of variables and assign local variables specific
scalar types where possible (LLVM uses a typed code representation).
The {\tt box} operations used to tag bit strings with types are done
lazily; they add a compile-time tag that causes generation of the
appropriate allocation code only when the value in question hits a context
that requires it (for example, assignment to an untyped data structure,
or being passed to an unknown function).

The code generator recognizes calls to key built-in and intrinsic functions,
and replaces them with efficient in-line code where possible. For example,
the {\tt is} function on mutable arguments yields a pointer comparison, and
{\tt typeof} might
yield a constant pointer value if the type of its argument is known.
Calls known to match single methods generate code to call the correct
method directly, skipping the dispatch process.

Finally, we run several of LLVM's optimization passes.
This provides standard scalar optimizations, such as
strength reduction, dead code elimination, jump threading, and constant
folding.

\section{Example Use Cases}

\subsection{Numeric Type Promotion}

Numeric types and arithmetic are fundamental to all programming, but deserve
extra attention in the case of scientific computing.
In traditional compiled languages such as C, the arithmetic operators are the
most polymorphic ``functions'', and hence cannot be written in the language
itself. Arithmetic must be defined in the compiler, including contentious
decisions such as how to handle operations with mixed argument types.

In Julia, multiple dispatch is used to define arithmetic and type
promotion behaviors at the library level rather than in the compiler.
As a result, the system smoothly incorporates new
operators and numeric types with minimal work.

Four key utility functions comprise the type promotion system.
For simplicity, we consider only two-argument forms of promotion
although multi-argument promotion is also defined and used.

\begin{enumerate}
\item {\tt convert(T, value)} converts its second argument to type {\tt T}
\item {\tt promote\_rule(T1,T2) } defines which of two types is greater in
the promotion partial order
\item {\tt promote\_type(T1,T2) } uses {\tt promote\_rule} to determine which
type should be used for values of types {\tt T1} and {\tt T2}
\item {\tt promote(v1, v2) } converts its arguments to an appropriate type
and returns the results
\end{enumerate}

{\tt promote} is implemented as follows:

\begin{verbatim}
function promote{T,S}(x::T, y::S)
    (convert(promote_type(T,S),x),
     convert(promote_type(T,S),y))
end
\end{verbatim}

{\tt promote\_type} simply tries {\tt promote\_rule} with its arguments in
both orders, to avoid the need for repeated definitions:

\begin{verbatim}
function promote_type{T,S}(::Type{T}, ::Type{S})
    if applicable(promote_rule, T, S)
        return promote_rule(T,S)
    elseif applicable(promote_rule, S, T)
        return promote_rule(S,T)
    else
        error("no promotion exists")
    end
end
\end{verbatim}

{\tt convert} and {\tt promote\_rule} are implemented for each type. Two
such definitions for the {\tt Complex128} type are:

\begin{verbatim}
promote_rule(::Type{Complex128},
             ::Type{Float64}) = Complex128
convert(::Type{Complex128}, x::Real) =
        complex128(x, 0)
\end{verbatim}

With these definitions in place, a function may gain generic promotion
behavior by adding the following kind of definition:

\begin{verbatim}
+(x::Number, y::Number) = +(promote(x,y)...)
\end{verbatim}

This means that, given two numeric arguments where no more specific
definition matches, promote the arguments and retry the operation
(the {\tt ...} ``splices'' the two values returned by {\tt promote} into
the argument list).
The standard library contains such definitions for all basic arithmetic
operators.
For this recursion to terminate, we require only that each {\tt Number}
type implement {\tt +} for two arguments of that type, e.g.

\begin{verbatim}
+(x::Int64, y::Int64) = ...
+(x::Float64, y::Float64) = ...
+(x::Complex128, y::Complex128) = ...
\end{verbatim}

Therefore, each new type requires only one definition of each operator,
and a handful of {\tt convert} and {\tt promote\_rule} definitions.
If $n$ is the number of types and $m$ is the number of operators, a new
type requires $O(n+m)$ rather than $O(n\cdot m)$ definitions.

The reader will notice that uses of this mechanism involve multiple method
calls, as well as potentially expensive features such as tuple allocation
and argument splicing. Without a sufficient optimizing compiler, this
implementation would be completely impractical. Fortunately, through
type analysis, inlining, elision of unnecessary tuples, and lowering of
the {\tt apply} operation implied by {\tt ...}, Julia's compiler is able
to eliminate all of the overhead in most cases, ultimately yielding a
sequence of machine instructions comparable to that emitted by a
traditional compiler.

The most troublesome function is {\tt promote\_type}. For good performance,
we must elide calls to it, but doing so may be incorrect since the function
might throw an error. By fortunate coincidence though, the logic in
{\tt promote\_type} exactly mirrors the analysis done by type inference: it
only throws an error if no matching methods exist for its calls to
{\tt promote\_rule}, in which case type inference concludes that the
function throws an error regardless of which branch is taken.
{\tt applicable} is a built-in function known to be free of effects.
Therefore, whenever a sharp result type for {\tt promote\_type} can be
inferred, it is also valid to remove the unused arms of the conditional.

\subsection{Code Generation and Staged Functions}

The presence of types and an inference pass creates a new, intermediate
translation stage which may be customized (macros essentially customize
syntax processing, and object systems customize run time behavior).
This is the stage at which types are known, and it exists in Julia via
the compiler's method specialization machinery. Specialization may occur
at run time during dispatch, or at compile time when inference is able
to determine argument types accurately.
Running custom code at this stage has two tremendous effects:
first, optimized code can be generated for special cases, and
second, the type inference system can effectively be extended to be able to
make new type deductions relevant to the user's application.

\begin{figure*}[t]
\begin{center}
\includegraphics[width=6in]{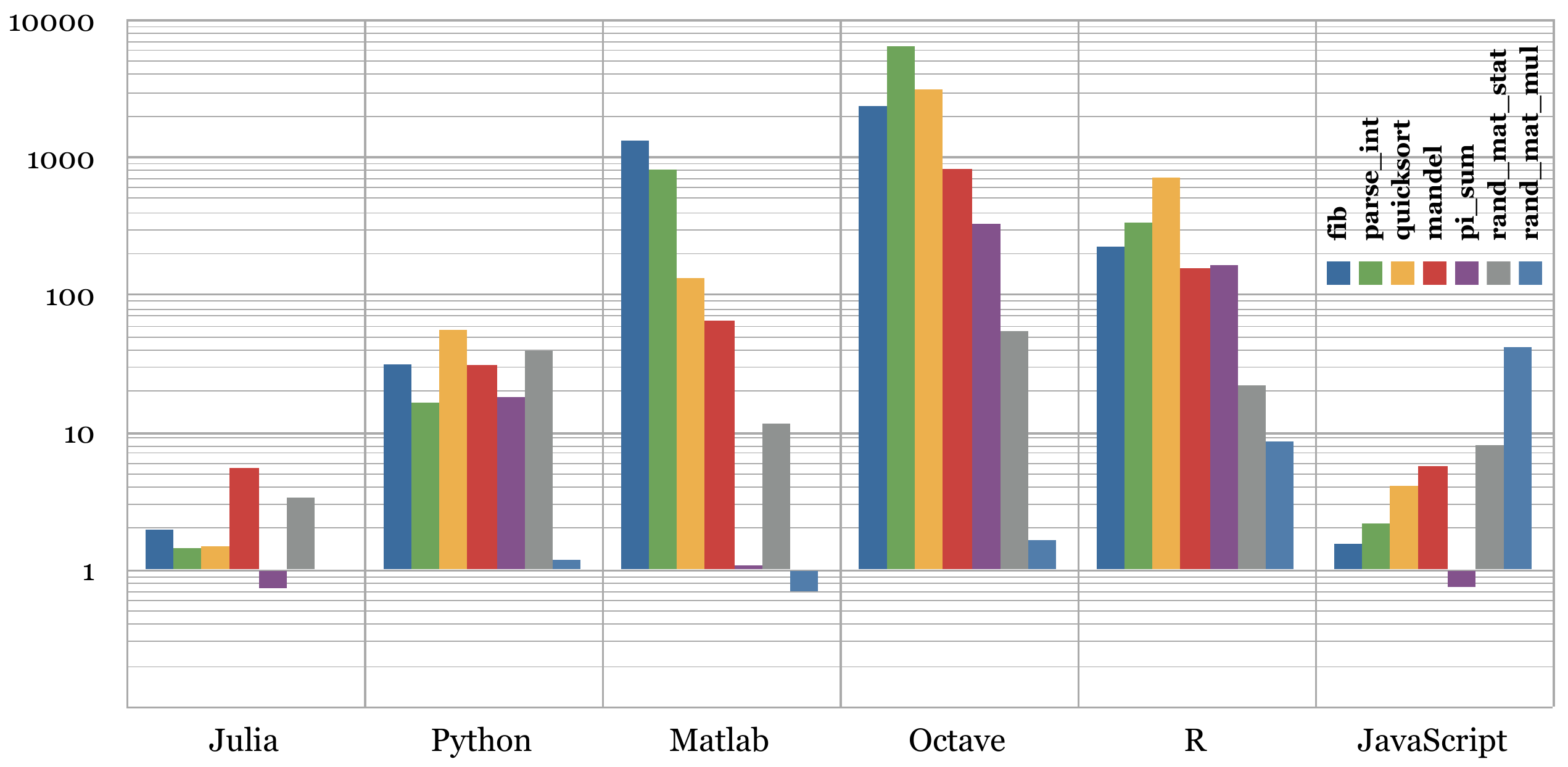}
\end{center}
\caption{
Microbenchmark results (times relative to C++, log-scale). These
measurements were carried out on a MacBook Pro with a 2.53GHz Intel Core 2
Duo CPU and 8GB of 1066MHz DDR3 RAM. The following versions were used:
Python 2.7.1, {\Matlab} R2011a, Octave 3.4, R 2.14.2, V8 3.6.6.11. The C++
baseline was compiled by GCC 4.2.1, taking best timing from all
optimization levels. Native implementations of array operations, matrix
multiplication, sorting, are used where available.
}
\label{mbr}
\end{figure*}

For example, we might want to write functions that apply to two
arrays of different dimensionality, where the result has the higher of the
two argument dimensionalities. One such function is a ``broadcasting''
binary elementwise operator, that performs computations such as adding a
column vector to every column of a matrix, or adding a plane to every slice
of a 3-dimensional dataset. We can determine the shape of the result
array with the following function:

\begin{verbatim}
function promote_shape(s1::Tuple, s2::Tuple)
    if length(s1) > length(s2)
        return s1
    else
        return s2
    end
end
\end{verbatim}

The type system can easily express the types of array shapes, for example
{\tt (Int,Int)} and {\tt (Int,Int,Int)}. However, inferring a sharp result
type for this simple function is still challenging. The inference algorithm
would have to possess a theory of the {\tt length} and {\tt >} functions,
which is not easily done given that all Julia functions may be redefined
and overloaded with arbitrary methods.

Instead, this function can be written as a \emph{staged function} (or
more accurately in our case, a \emph{staged method}). This is a function
that runs at an earlier translation ``stage'', i.e. compile time, and
instead of returning a result value returns code that will compute the
result value when executed \cite{staging}.
Here is the staged version of
{\tt promote\_shape}\footnote{The {\tt @} denotes a macro invocation. At
present, staged methods are implemented by a macro, but full integration
into the language is planned.}:

\begin{verbatim}
@staged function promote_shape(s1::Tuple, s2::Tuple)
    if length(s1) > length(s2)
        quote return s1 end
    else
        quote return s2 end
    end
end
\end{verbatim}

The signature of this definition behaves exactly like any other method
signature: the type annotations denote run-time types for which the
definition is applicable. However, the body of the method will be invoked
on the \emph{types} of the arguments rather than actual arguments, and the
result of the body will be used to generate a new, more specialized
definition. For example, given arguments of types
{\tt (Int,Int)} and {\tt (Int,Int,Int)} the generated definition would be:

\begin{verbatim}
function promote_shape(s1::(Int,Int),
                       s2::(Int,Int,Int))
    return s2
end
\end{verbatim}

Observe that the type of this function is trivial to infer.

The staged function body runs as normal user code, so whatever definition
of {\tt >} is visible will be used, and the compiler does not have to know
how it behaves. Critically, the staged version of the function looks
similar to the normal version, requiring only the insertion of {\tt quote}
to mark expressions deferred to the next stage.

In the case where a program is already statically-typeable, staged
functions preserve that property. The types of the arguments to the
staged function will be known at compile time, so the custom code
generator can be invoked at compile time. Then the compiler may inline
the result or emit a direct call to the generated code, as usual.

Or, if the user does not require static compilation, the custom code
generator can be invoked at run time. Its results are cached for each new
combination of argument types, so compilation pauses are infrequent.

As a result, functions with complex type behavior can be implemented
in libraries without losing performance. Of course, ordinary Julia
functions may also have complex type behavior, and it is up to the
library designer to decide which functions should be staged.

\section{Evaluation}

To evaluate Julia's performance, we have compared its speed to that of six
other languages: C++, Python, \Matlab, Octave, R, and JavaScript.
Figure~\ref{mbr}
shows timings for five scalar microbenchmarks, and two
simple array benchmarks; the same data are presented in tabular format in Table~\ref{mbrtab}. All numbers are ratios relative to
the time taken by C++. The first five tests do not
reflect typical application performance in each environment; their only
purpose is to compare the code generation and execution for basic language
constructs, such as manipulating scalar quantities and referencing individual
array elements.

\begin{table*}[t]
\begin{center}
\begin{tabular}{|l|r|r|r|r|r|r|}\hline
test & Julia & Python & \Matlab & Octave & R & JavaScript \\
\hline \hline
fib        & 1.97 & 31.47 & 1336.37  & 2383.80 & 225.23 & 1.55 \\
\hline
parse\_int & 1.44 & 16.50 &  815.19  & 6454.50 & 337.52 & 2.17 \\
\hline
quicksort  & 1.49 & 55.84 &  132.71  & 3127.50 & 713.77 & 4.11 \\
\hline
mandel     & 5.55 & 31.15 &   65.44  &  824.68 & 156.68 & 5.67 \\
\hline
pi\_sum    & 0.74 & 18.03 &    1.08  &  328.33 & 164.69 & 0.75 \\
\hline
rand\_mat\_stat & 3.37 & 39.34 & 11.64 & 54.54 &  22.07 & 8.12 \\
\hline
rand\_mat\_mul  & 1.00 &  1.18 &  0.70 &  1.65 &   8.64 & 41.79 \\
\hline
\end{tabular}
\end{center}
\caption{Microbenchmark results (times relative to C++).}
\label{mbrtab}
\end{table*}

\Matlab has a JIT compiler
that works quite well in some cases, but is inconsistent, and
performs especially poorly on user-level function calls. The V8
JavaScript JIT compiler's performance is impressive. Anomalously, both
Julia and JavaScript seem to beat C++ on pi\_sum, but we have not yet
discovered why this might be.

The rand\_mat\_stat code manipulates many 5-by-5 matrices. Here the
performance gaps close, but the arrays are not large enough for
library time to dominate, so Julia's ability to specialize call sites
wins the day (despite the fact that most of the array library functions
involved are written in Julia itself).

The rand\_mat\_mul code demonstrates a case where time spent in BLAS \cite{blas}
dominates. \Matlab gets its edge from using a
multi-threaded BLAS (threading is available in the BLAS Julia uses,
but it was disabled when these numbers were taken). R may not be using
a well-tuned BLAS in this install; more efficient configurations are
probably possible.
JavaScript as typically deployed is not able to call the native BLAS code,
but the V8 compiler's work is respectable here.

Julia is not yet able to cache generated native code, and so incurs a
startup time of about two seconds to compile basic library functions.
For some applications this latency is a barrier to deployment, and we plan
to address it in the future.

\subsection{Effectiveness of Specialization Heuristics}

Given our implementation strategy, excessive compilation and corresponding
memory use are potential performance concerns.
We measured the number of method compilations performed both with and
without our specialization heuristics, and
the heuristics were able to elide about 12\% of compilations. This is
not a large fraction, but it is satisfying given that the heuristics can
be computed easily, and only by manipulating types. On average, each
method is compiled about 2.5 times.


Memory usage is not unreasonable for modern machines: on a 64-bit platform
Julia uses about 50MB of memory on startup, and after loading several
libraries and working for a while memory use tends to level off around
150-200MB. Pointer-heavy data structures consume a lot of space on
64-bit platforms. To mitigate this problem, we store ASTs and type
information in a compact serialized format, and deserialize structures
when the compiler needs them.

\subsection{Effectiveness of Type Inference}

It is interesting to count compiled expressions for which
a concrete type can be inferred. In some sense, this tells us ``how close''
Julia is to being statically typed, though in our case this is a property
of both the language implementation and the standard library.
In a run of our test suite, code was generated for 135375 expressions.
Of those, 84127 (62\%) had a type more specific than {\tt Any}. Of those,
80874 (96\%) had a concrete static type.

This suggests that use of dynamic typing is fairly popular, even though
we try to avoid it to some extent in the standard library. Still, more
than half of our code is well-typed. The numbers also suggest that,
despite careful use of a rich lattice, typing tends to be an all-or-nothing
affair. But, it is difficult to estimate the effect of the 4\%
abstractly-typed expressions on the other 96\%, not to mention the potential
utility of abstract inferred types in code that was not actually
compiled.

These numbers are somewhat inaccurate, as they include dead code, and
it may be the case that better-typed methods tend to be recompiled with
different frequency than others, biasing the numbers.

\subsection{Productivity}

Our implementation of Julia consists of 11000 lines of C, 4000 lines
of C++, and 3500 lines of Scheme (here we are not counting code in
external libraries such as BLAS and LAPACK).  Thus we have
significantly less low-level code to maintain than most scripting
languages.  Our standard library is roughly 25000 lines of Julia code.
The standard library provides around 300 numerical functions of the
sort found in all technical computing environments. We suspect that
our library is one of the most compact implementations of this body of
functionality.

At this time, every contributor except the core developers is a ``new
user'' of Julia, having known of the language for no more than six
months.  Despite this, our function library has received several
significant community contributions, and numerous smaller ones. We
take this as encouraging evidence that Julia is productive and easy to
learn.

\section{Community}

Julia is an open source project, with all code hosted on
{\tt github} \cite{github}.
It has attracted 550 mailing list subscribers, 1500 github followers,
190 forks, and more than 50 total contributors. Text editor support
has been implemented for emacs, vim, and textmate.
Github recognizes Julia as the language of source files ending in
{\tt .jl}, and can syntax highlight Julia code listings.

Several community projects are underway: two plotting packages,
interfaces to arbitrary-precision arithmetic library GMP,
bit arrays, linear programming, image processing, polynomials,
GPU code generation, a statistics library, and a web-based interactive
environment. A package management framework will soon be in place.

We hope Julia is part of a new generation of dynamic languages that not
only run faster, but foster more cooperation between the programmer
and compiler, pushing the standard of productivity ever higher.




\section{Acknowledgements}
We wish to thank our funding agencies for their support via
Department of Energy grant DE-SC0002099 and National Science Foundation
grant CCF-0832997. We also gratefully acknowledge gifts from
VMWare and Citigroup.


\bibliographystyle{abbrv}

\begin{thebibliography}{29}

\providecommand{\natexlab}[1]{#1}
\providecommand{\url}[1]{\texttt{#1}}
\expandafter\ifx\csname urlstyle\endcsname\relax
  \providecommand{\doi}[1]{doi: #1}\else
  \providecommand{\doi}{doi: \begingroup \urlstyle{rm}\Url}\fi

\bibitem[Allen et~al.(2011)Allen, Hilburn, Kilpatrick, Luchangco, Ryu, Chase,
  and Steele]{modularmultipledispatch}
E.~Allen, J.~Hilburn, S.~Kilpatrick, V.~Luchangco, S.~Ryu, D.~Chase, and
  G.~Steele.
\newblock Type checking modular multiple dispatch with parametric polymorphism
  and multiple inheritance.
\newblock In \emph{Proceedings of the 2011 ACM international conference on
  Object oriented programming systems languages and applications}, OOPSLA '11,
  pages 973--992, New York, NY, USA, 2011. ACM.

\bibitem[Anderson et~al.(2005)Anderson, Giannini, and
  Drossopoulou]{typeinfjavascript}
C.~Anderson, P.~Giannini, and S.~Drossopoulou.
\newblock Towards type inference for javascript.
\newblock In A.~Black, editor, \emph{ECOOP 2005 - Object-Oriented Programming},
  volume 3586 of \emph{Lecture Notes in Computer Science}, pages 733--733.
  Springer Berlin / Heidelberg, 2005.

\bibitem[Baker(1990)]{nimble}
H.~G. Baker.
\newblock The nimble type inferencer for common lisp-84.
\newblock Technical report, Tech. Rept., Nimble Comp, 1990.

\bibitem[Baker(1991)]{clostrophobia}
H.~G. Baker.
\newblock Clostrophobia: its etiology and treatment.
\newblock \emph{SIGPLAN OOPS Mess.}, 2\penalty0 (4):\penalty0 4--15, Oct. 1991.

\bibitem[Beer(1987)]{pticl}
R.~D. Beer.
\newblock Preliminary report on a practical type inference system for common
  lisp.
\newblock \emph{SIGPLAN Lisp Pointers}, 1:\penalty0 5--11, June 1987.

\bibitem[Bolz et~al.(2009)Bolz, Cuni, Fijalkowski, and Rigo]{pypyjit}
C.~F. Bolz, A.~Cuni, M.~Fijalkowski, and A.~Rigo.
\newblock Tracing the meta-level: Pypy's tracing jit compiler.
\newblock In \emph{Proceedings of the 4th workshop on the Implementation,
  Compilation, Optimization of Object-Oriented Languages and Programming
  Systems}, ICOOOLPS '09, pages 18--25, New York, NY, USA, 2009. ACM.

\bibitem[Brooks and Gabriel(1984)]{lispcrit}
R.~A. Brooks and R.~P. Gabriel.
\newblock A critique of common lisp.
\newblock In \emph{Proceedings of the 1984 ACM Symposium on LISP and functional
  programming}, LFP '84, pages 1--8, New York, NY, USA, 1984. ACM.

\bibitem[Cardelli and Wegner(1985)]{boundedquant}
L.~Cardelli and P.~Wegner.
\newblock On understanding types, data abstraction, and polymorphism.
\newblock \emph{ACM Comput. Surv.}, 17\penalty0 (4):\penalty0 471--523, Dec.
  1985.

\bibitem[Chambers(1992)]{cecil}
C.~Chambers.
\newblock Object-oriented multi-methods in cecil.
\newblock In \emph{Proceedings of the European Conference on Object-Oriented
  Programming}, pages 33--56, London, UK, 1992. Springer-Verlag.

\bibitem[Chambers(2005)]{dieselspec}
C.~Chambers.
\newblock The diesel language specification and rationale: Version 0.1.
\newblock February 2005.

\bibitem[Chambers et~al.(1989)Chambers, Ungar, and Lee]{selflang}
C.~Chambers, D.~Ungar, and E.~Lee.
\newblock An efficient implementation of self: a dynamically-typed
  object-oriented language based on prototypes.
\newblock \emph{SIGPLAN Not.}, 24:\penalty0 49--70, September 1989.

\bibitem[Cousot and Cousot(1977)]{abstractinterp}
P.~Cousot and R.~Cousot.
\newblock Abstract interpretation: a unified lattice model for static analysis
  of programs by construction or approximation of fixpoints.
\newblock In \emph{Proceedings of the 4th ACM SIGACT-SIGPLAN symposium on
  Principles of programming languages}, POPL '77, pages 238--252, New York, NY,
  USA, 1977. ACM.

\bibitem[Cousot and Cousot(1992)]{widening}
P.~Cousot and R.~Cousot.
\newblock Comparing the galois connection and widening/narrowing approaches to
  abstract interpretation.
\newblock In M.~Bruynooghe and M.~Wirsing, editors, \emph{Programming Language
  Implementation and Logic Programming}, volume 631 of \emph{Lecture Notes in
  Computer Science}, pages 269--295. Springer Berlin / Heidelberg, 1992.

\bibitem[Dabbish et~al.(2012)Dabbish, Stuart, Tsay, and Herbsleb]{github}
L.~Dabbish, C.~Stuart, J.~Tsay, and J.~Herbsleb.
\newblock Social coding in github: transparency and collaboration in an open
  software repository.
\newblock In \emph{Proceedings of the ACM 2012 conference on Computer Supported
  Cooperative Work}, CSCW '12, pages 1277--1286, New York, NY, USA, 2012. ACM.

\bibitem[DeMichiel and Gabriel(1987)]{closoverview}
L.~DeMichiel and R.~Gabriel.
\newblock The common lisp object system: An overview.
\newblock In J.~Bézivin, J.-M. Hullot, P.~Cointe, and H.~Lieberman, editors,
  \emph{ECOOP’ 87 European Conference on Object-Oriented Programming}, volume
  276 of \emph{Lecture Notes in Computer Science}, pages 151--170. Springer
  Berlin / Heidelberg, 1987.

\bibitem[Furr et~al.(2009)Furr, An, and Foster]{profileguided}
M.~Furr, J.-h.~D. An, and J.~S. Foster.
\newblock Profile-guided static typing for dynamic scripting languages.
\newblock \emph{SIGPLAN Not.}, 44:\penalty0 283--300, Oct. 2009.

\bibitem[Gomez(1999)]{scilab}
C.~Gomez, editor.
\newblock \emph{Engineering and Scientific Computing With Scilab}.
\newblock Birkh{\"a}user, 1999.

\bibitem[Ihaka and Gentleman(1996)]{Rlang}
R.~Ihaka and R.~Gentleman.
\newblock R: A language for data analysis and graphics.
\newblock \emph{Journal of Computational and Graphical Statistics}, 5:\penalty0
  299--314, 1996.

\bibitem[J{\o}rring and Scherlis(1986)]{staging}
U.~J{\o}rring and W.~L. Scherlis.
\newblock Compilers and staging transformations.
\newblock In \emph{Proceedings of the 13th ACM SIGACT-SIGPLAN symposium on
  Principles of programming languages}, POPL '86, pages 86--96, New York, NY,
  USA, 1986. ACM.

\bibitem[Kaplan and Ullman(1980)]{kaplanullman}
M.~A. Kaplan and J.~D. Ullman.
\newblock A scheme for the automatic inference of variable types.
\newblock \emph{J. ACM}, 27:\penalty0 128--145, January 1980.

\bibitem[Lattner and Adve(2004)]{LLVM}
C.~Lattner and V.~Adve.
\newblock {LLVM: A Compilation Framework for Lifelong Program Analysis \&
  Transformation}.
\newblock In \emph{{Proceedings of the 2004 International Symposium on Code
  Generation and Optimization (CGO'04)}}, Palo Alto, California, Mar 2004.

\bibitem[Lawson et~al.(1979)Lawson, Hanson, Kincaid, and Krogh]{blas}
C.~L. Lawson, R.~J. Hanson, D.~R. Kincaid, and F.~T. Krogh.
\newblock Basic linear algebra subprograms for fortran usage.
\newblock \emph{ACM Trans. Math. Softw.}, 5\penalty0 (3):\penalty0 308--323,
  Sept. 1979.

\bibitem[Ma and Kessler(1990)]{TICL}
K.-L. Ma and R.~R. Kessler.
\newblock Ticl—a type inference system for common lisp.
\newblock \emph{Software: Practice and Experience}, 20\penalty0 (6):\penalty0
  593--623, 1990.

\bibitem[Mohnen(2002)]{graphfree}
M.~Mohnen.
\newblock A graph—free approach to data—flow analysis.
\newblock In R.~Horspool, editor, \emph{Compiler Construction}, volume 2304 of
  \emph{Lecture Notes in Computer Science}, pages 185--213. Springer Berlin /
  Heidelberg, 2002.

\bibitem[Morandat et~al.(2012)Morandat, Hill, Osvald, and Vitek]{evaluatingR}
F.~Morandat, B.~Hill, L.~Osvald, and J.~Vitek.
\newblock Evaluating the design of the R language.
\newblock In J.~Noble, editor, \emph{ECOOP 2012 – Object-Oriented
  Programming}, volume 7313 of \emph{Lecture Notes in Computer Science}, pages
  104--131. Springer Berlin / Heidelberg, 2012.

\bibitem[Murphy(1997)]{Octave}
M.~Murphy.
\newblock Octave: A free, high-level language for mathematics.
\newblock \emph{Linux J.}, 1997, July 1997.

\bibitem[Robin and Milner(1978)]{MLtypeinf}
Robin and Milner.
\newblock A theory of type polymorphism in programming.
\newblock \emph{Journal of Computer and System Sciences}, 17\penalty0
  (3):\penalty0 348 -- 375, 1978.

\bibitem[Shalit(1996)]{dylanlang}
A.~Shalit.
\newblock \emph{The Dylan reference manual: the definitive guide to the new
  object-oriented dynamic language}.
\newblock Addison Wesley Longman Publishing Co., Inc., Redwood City, CA, USA,
  1996.

\bibitem[van~der Walt et~al.(2011)van~der Walt, Colbert, and Varoquaux]{numpy}
S.~van~der Walt, S.~C. Colbert, and G.~Varoquaux.
\newblock The numpy array: a structure for efficient numerical computation.
\newblock \emph{CoRR}, abs/1102.1523, 2011.

\end{thebibliography}


\end{document}